\documentclass[]{spie}  

\usepackage{amsmath,amsfonts,amssymb}
\usepackage{graphicx}
\usepackage[colorlinks=true, allcolors=blue]{hyperref}

\newcommand{\V}[1]{\boldsymbol{#1}} 
\newcommand{\Vx}{\V{x}} 

\newcommand{\GP}[1]{\left(#1\right)}
\newcommand{\GC}[1]{\left[#1\right]}

\newcommand{\EST}[1]{\widehat{#1}}

\title{FIRST, a Pupil-Remapping Fiber Interferometer at the Subaru Telescope: on-sky results}

\author[a,b]{S. Vievard}
\author[c]{E. Huby}
\author[c]{S. Lacour}
\author[c]{K. Barjot}
\author[d]{G. Martin}
\author[e]{N. Cvetojevic}
\author[a,c]{V. Deo}
\author[a,b,f]{O. Guyon}
\author[a]{J. Lozi}
\author[b]{T. Kotani}
\author[h]{N. Jovanovic}
\author[i,c]{F. Marchis}
\author[d,j]{G. Duchène}
\author[c]{V. Lapeyrere}
\author[c]{D. Rouan}
\author[c]{G. Perrin}

\affil[a]{National Astronomical Observatory of Japan, Subaru Telescope, 650 North Aohoku Place,
Hilo, HI 96720, U.S.A.}
\affil[b]{Astrobiology Center of NINS, 2-21-1, Osawa, Mitaka, Tokyo, 181-8588, Japan}
\affil[c]{LESIA, Observatoire de Paris, Université PSL, CNRS, Sorbonne Université, Université de Paris, 5 place Jules Janssen, 92195 Meudon, France}
\affil[d]{Univ. Grenoble Alpes, CNRS, IPAG, 38000 Grenoble, France}
\affil[e]{Observatoire de la C\^ote d'Azur, 96 Boulevard de l'Observatoire, 06300 Nice, France}
\affil[f]{College of Optical Sciences, University of Arizona, Tucson, AZ 85721, U.S.A.}
\affil[h]{California Institute of Technology, 1200 E California Blvd, Pasadena, CA 91125, U.S.A.}
\affil[i]{Carl Sagan Center at the SETI Institute, 189 Bernardo Av., Mountain View, CA 94043, USA}
\affil[j]{Department of Astronomy, University of California at Berkeley, Berkeley, CA 94720-3411}

\authorinfo{Further author information: Sebastien Vievard: E-mail: vievard@naoj.org}

\begin{document} 
\maketitle

\begin{abstract}
FIRST, the Fibered Imager foR a Single Telescope, is a spectro-imager using single-mode fibers for pupil remapping, allowing measurements beyond the telescope diffraction limit. Integrated on the Subaru Coronagraphic Extreme Adaptive Optics instrument at the Subaru Telescope, it benefits from a very stable visible light wavefront allowing to acquire long exposure and operate on significantly fainter sources than previously possible. On-sky results demonstrated the ability of the instrument to detect stellar companions separated 43mas in the case of the Capella binary system. A similar approach on an extremely large telescope would offer unique scientific opportunities for companion detection and characterization at very high angular resolution.

\end{abstract}


\section{INTRODUCTION}
\label{sec:intro}  

One of the key challenges in astronomy is the study of circumstellar environment, \textit{e.g.} the detection and characterization of stellar companions. Achieving this requires both high performance in terms of angular resolution, provided by the size of the telescope, and dynamic range, provided by high contrast instruments and data processing. Even though very large telescopes (8-meter class) were built and Extremely Large Telescopes (ELTs, 30-meter class) are on the horizon, reaching the desired performances is quite challenging especially for ground based telescopes, subject to turbulent atmosphere. 

To try to use the telescopes at their full potential, several techniques were developed. Among them, the Adaptive Optics (AO) technique~\cite{rousset1990first} consists in using a deformable mirror to correct for the turbulence-induce aberrations on the incoming wavefront, allowing to restore the diffraction limit of the telescope. However, because this correction is not perfect, and because of other instrumental limitations like Non-Common Path Aberrations, residual speckle noise still limits the achievable dynamic range. Other techniques were investigated. One of them, based on the post-processing of short exposure images, is called Speckle interferometry~\cite{labeyrie1970attainment} . It allows to recover the diffraction limit of the telescope, but suffers from a contrast limitation around 100. Another technique, called aperture masking, consists in placing a non-redundant mask in the pupil of the telescope and recombine the coherent light from each sub-aperture. This allows to retrieve spatial information at the highest spatial frequency of the telescope with contrast ratios down to $10^{-3}$~\cite{lacour2011sparse}. However a rather large percentage of the pupil light collection needs to be sacrificed because of the non-redundancy needed for the mask. In addition, speckle noise remains across each sub-aperture.

The method that can make use of the aperture masking concept without its disadvantages is the pupil remapping technique. It consists in sampling the pupil and, with the help of single-mode fibers, recombining the different sub-pupils\cite{perrin2006high} . This allows 1- to exploit the light from the whole pupil, and 2- to "clean" the wavefront from each sub-pupil (therefore, remove the speckle noise) thanks to the natural spatial filtering ability provided by the single-mode fibers. Finally, such a concept has the capability to retrieve spatial information about an observed astronomical object below the telescope theoretical diffraction limit. An instrument with such features is called FIRST (Fibered Imager foR a Single Telescope) and is currently installed on the 8-meter Subaru Telescope as part of the Subaru Coronagraphic Extreme Adaptive Optics (SCExAO) instrument~\cite{2015PASP..127..890J,vievard2020capabilities} . Prior to its deployment on the Subaru Telescope, FIRST revealed its power on the 3-meters Lick observatory performing spectroscopy analysis of the Capella binary system, at the telescope's diffraction limit~\cite{huby2012first,huby2013first} . 

We present hereafter the principle of FIRST and its integration as a module of SCExAO. We then present the data reduction process used to extract the information from the recombined sub-pupils. Finally, as a first on-sky demonstration on the Subaru Telescope, we present the detection of the Capella system.

\section{Pupil remapping interferometer at the Subaru Telescope}

\subsection{Principle}

\begin{figure}[!h]
    \centering
    \includegraphics[width=\linewidth]{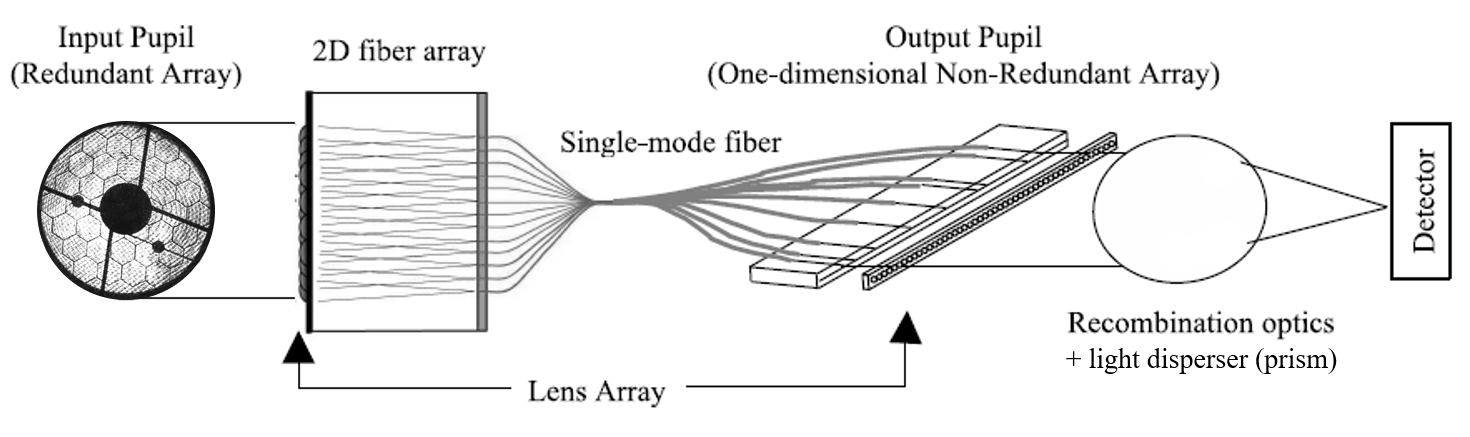}
    \caption{Schematic diagram~\cite{kotani2009pupil} showing the principle of the FIRST instrument at the Subaru Telescope. The pupil of the Subaru Telescope is divided into several single-mode fibers thanks to a 2D micro-lens array. Light from the different fibers are then set into a one-dimensional non-redundant array, recombined and spectrally dispersed.  }
    \label{FIRST_principle}
\end{figure}

The FIRST instrument elegantly combines the aperture masking technique with the use of single mode fibers to do pupil remapping. The use of single mode fibers allows to 1- use a larger section of the pupil compared to a simple pupil mask with holes, 2- "clean" each sub-pupil wavefront by removing the speckle noise. The idea of the instrument, described in Fig~\ref{FIRST_principle}, is to divide the 2D telescope pupil into several single-mode fibers using a micro-lens array. The latter can then be re-arranged into a non-redundant configuration allowing to retrieve independently, after recombination, $N/(N-1)/2$ spatial frequencies from the original pupil ($N$ being the number of sub-pupils). The use of a light disperser can then allow to obtain phase information (from the interference pattern) as a function of the wavelength. 

{}

\subsection{Setup of the instrument}

\begin{figure}[!h]
    \centering
    \includegraphics[width=\linewidth]{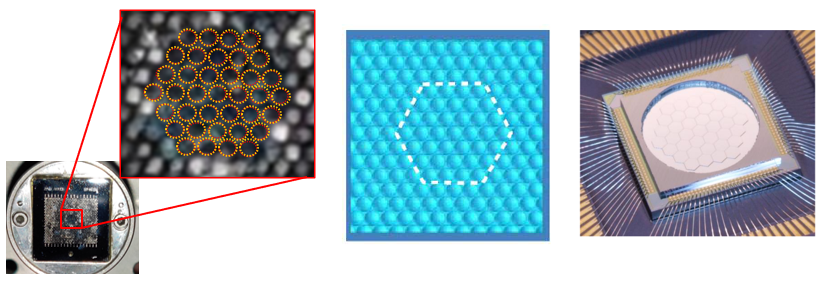}
    \caption{Left: Fiber bundle holding 36~fibers positioned on a hexagonal grid. Middle: 2D micro-lens array used to sample the pupil. Right: Segmented mirror, used to optimize the injection of each sub-pupil into the single-mode fibers.}
    \label{FIRST_hardware}
\end{figure}

The current FIRST instrument has the ability to sample the pupil into $18$~fibers (2 independent sets of $9$~fibers each). A fiber bundle holds $36$~single mode fibers together on a $250\mu m$-pitch hexagonal grid (see Fig.\ref{FIRST_hardware}-left). The input pupil is injected into the single mode fibers with a 2D micro-lens array (see Fig.~\ref{FIRST_hardware}-middle). The pitch of the latter is $250\mu m$ to match the fiber bundle and each lens focal length is $1mm$. The injection into the single-mode fibers is optimized thanks to a segmented mirror (Micro-ElectroMechanical System technology~-~MEMS) manufactured by Iris AO company (see Fig.~\ref{FIRST_hardware}-right). 

{}

For the recombination, the output of the single mode fibers are arranged in a V-groove. The output beams are collimated, then recombined, and spectrally dispersed with an equilateral SF2-prism. In order to increase the spectral resolution, an afocal anamorphic system, consisting in cylindrical lenses, stretches the beam in the dispersion direction and compresses it in the orthogonal direction.

\subsection{FIRST, a module on SCExAO}

\begin{figure}[!h]
    \centering
    \begin{tabular}{cc}
        \includegraphics[width=0.47\linewidth]{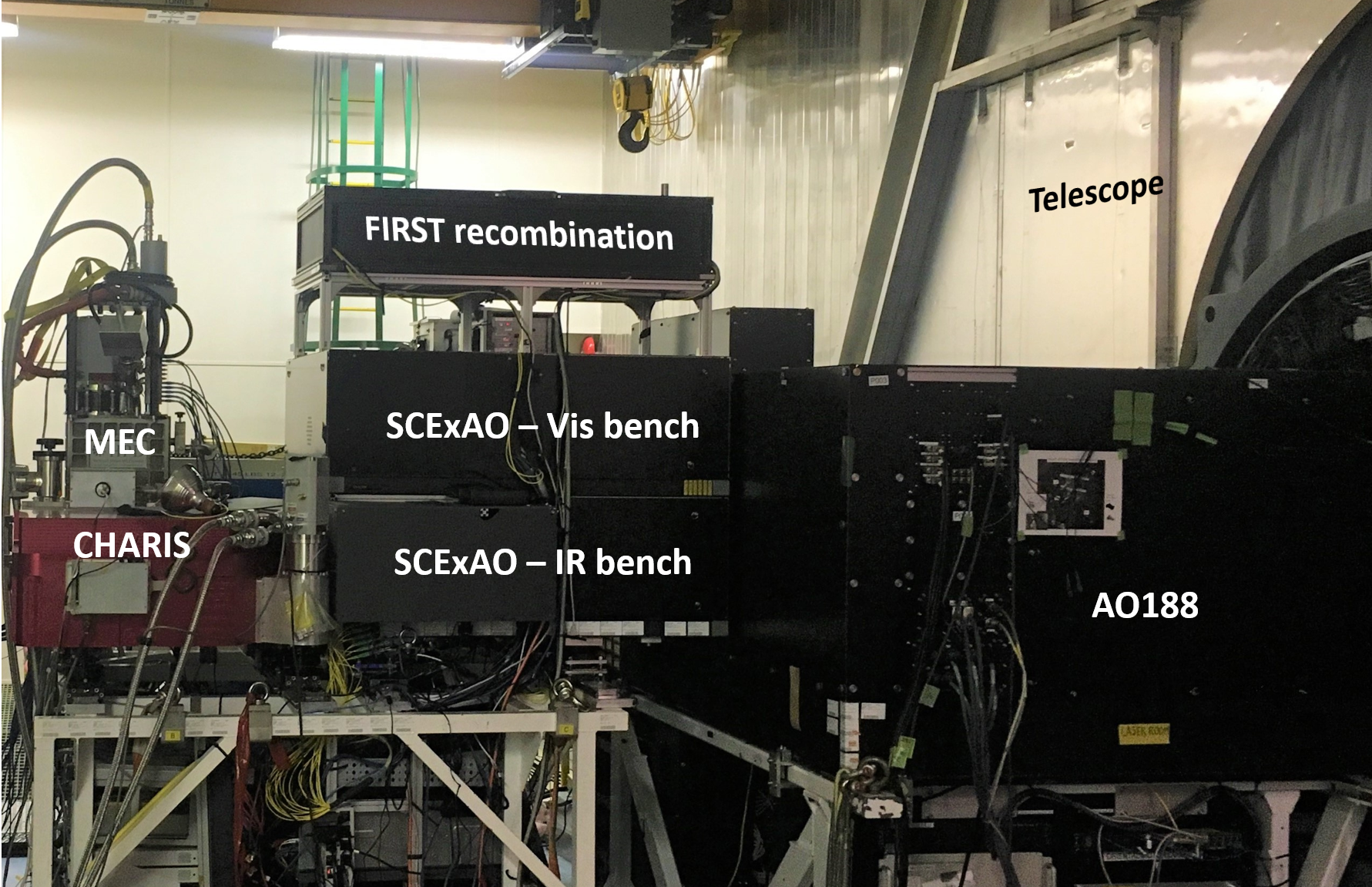} & \includegraphics[width=0.47\linewidth]{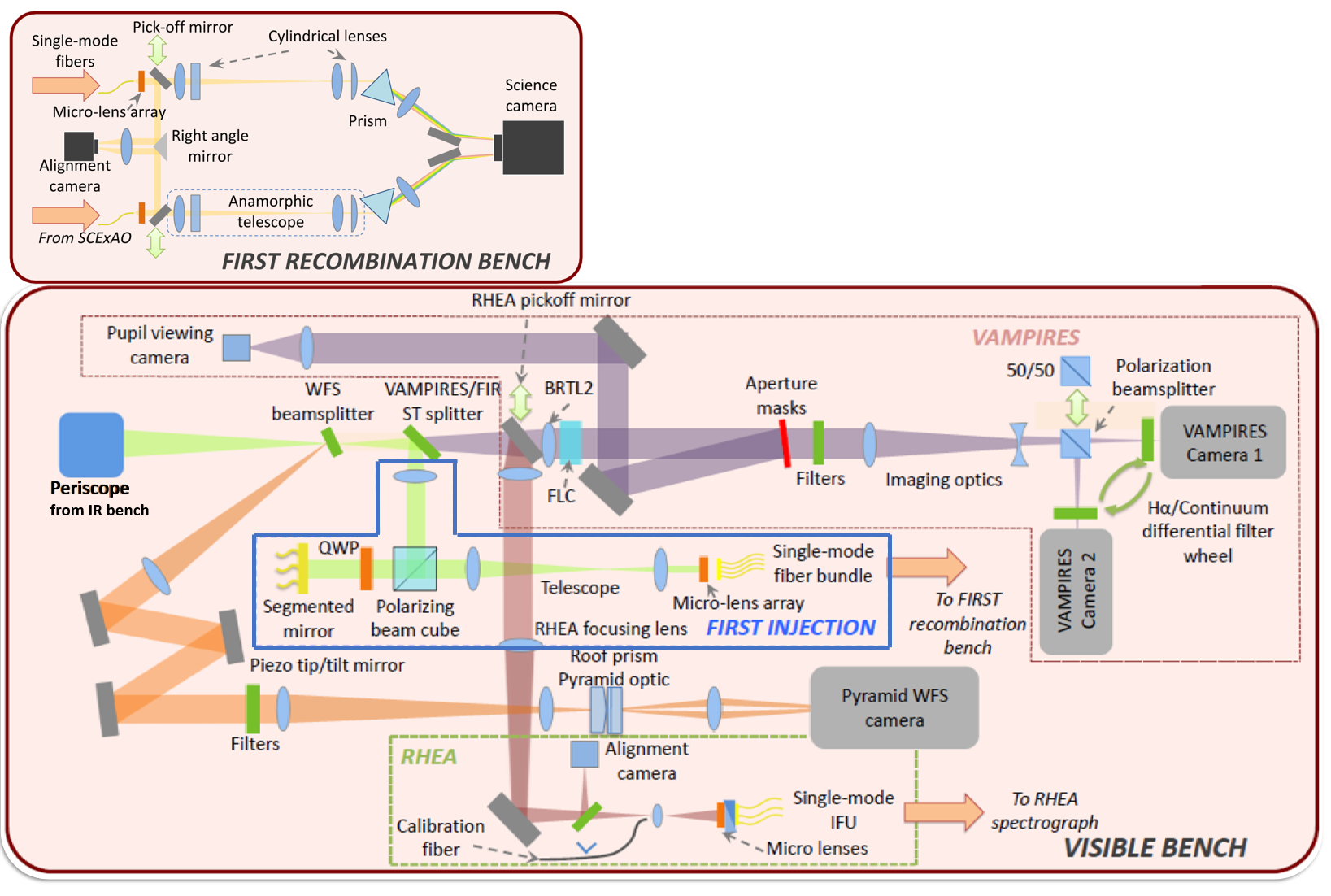} \\
    \end{tabular}{}
    \caption{Left: SCExAO on the Infra-Red (IR) Nasmyth platfrom sits after AO188. SCExAO is composed of two benches on the top of the other, plus the FIRST recombination bench. Right: A schematic of the optic path in the Visible bench and in the FIRST recombination bench.}
    \label{FIRST_scexao}
\end{figure}

SCExAO is a versatile platform dedicated to high contrast imaging on the Subaru Telescope. Located on the Nasmyth Infra-Red focus of the Subaru Telescope (see Fig.~\ref{FIRST_scexao}-left), it gets a partially corrected wavefront from the adaptive optics facility AO188~\cite{minowa2010performance}. In its essence, SCExAO is composed by 1- a pyramid wavefront sensor~\cite{Lozi_2019} operating in the Visible (around $800-950nm$) and delivering a wavefront quality over 80\% Strehl (in H-band) 2- several coronagraphs to suppress star light and reveal/study close circumstellar environment. Over the last 10 years SCExAO complexity increased and is now composed by 3~commissioned science modules~\cite{Currie_2018, vampires, jovanovic2017developing} and several experimental modules distributed over two optical benches: an IR ($950nm-2.5\mu m$) and a Visible ($600-950nm$) bench. 

{}

\begin{figure}[!h]
    \centering
    \includegraphics[width=0.7\linewidth]{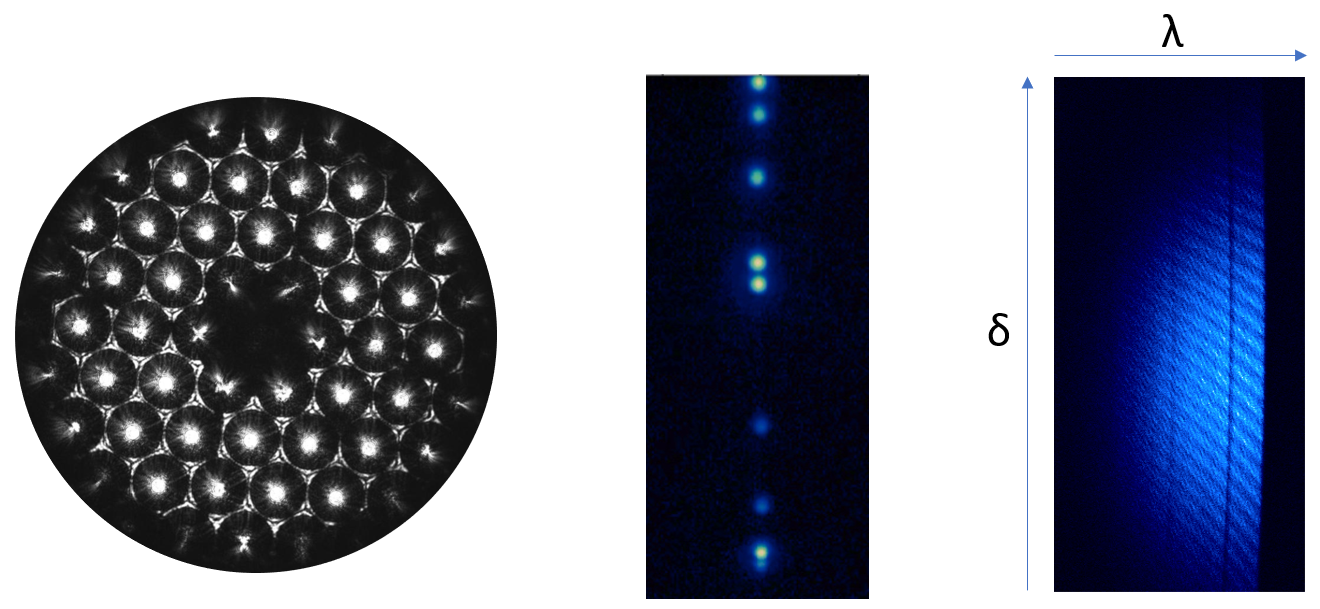}
    \caption{Left: Image of the pupil plan in FIRST, where the segments of the MEMS and the micro-lenses are conjugated. Middle: Image of the fibers outputs for photometric monitoring during the injection optimization. Right: Fringe pattern after recombination and spectral dispersion of the fiber outputs. }
    \label{FIRST_images}
\end{figure}

FIRST is divided in two different benches: the injection and the recombination. The injection part sits on the Visible bench (see Fig.~\ref{FIRST_scexao}-right). The Iris AO MEMS and the micro-lens array are both conjugated with the SCExAO pupil plane before injection into the single-mode fibers (see Fig.~\ref{FIRST_images}-left). The fibers of the bundle are connected to single mode fiber extensions (polished by hand for path length matching~\cite{huby2013caracterisation}) that run to the recombination bench located on the top of SCExAO (see Fig.~\ref{FIRST_scexao}-left). Inside the recombination bench, each of the fiber set can follow two different paths. One path is dedicated to the optimization of the injection. Here, the output of the fibers are imaged on a camera (see Fig.~\ref{FIRST_images}-middle) and an automatic procedure maximizes the flux by acting on the Iris AO MEMS segment tip/tilts. Once the injection is optimized, the beams from the fiber outputs go to the science path where the recombination and dispersing optics generate the fringe pattern on the science camera (see Fig.~\ref{FIRST_images}-right). The spectral resolution obtained is empirically around $300$ @$700$nm and FIRST field of view is about $136$mas at $700$nm.

{}

\begin{figure}[!h]
    \centering
    \begin{tabular}{cc}
        \includegraphics[width=0.43\linewidth]{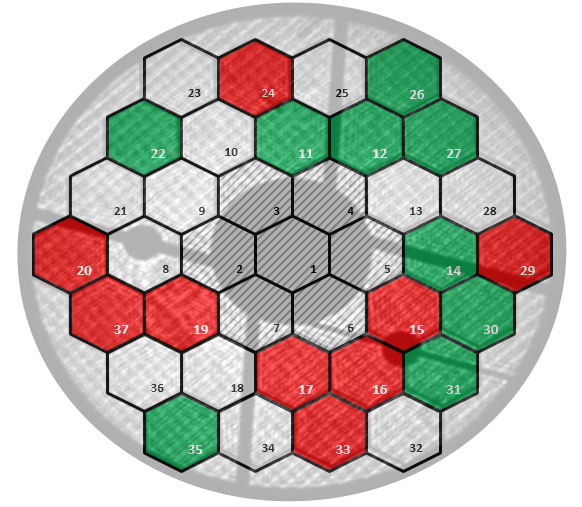} & \includegraphics[width=0.4\linewidth]{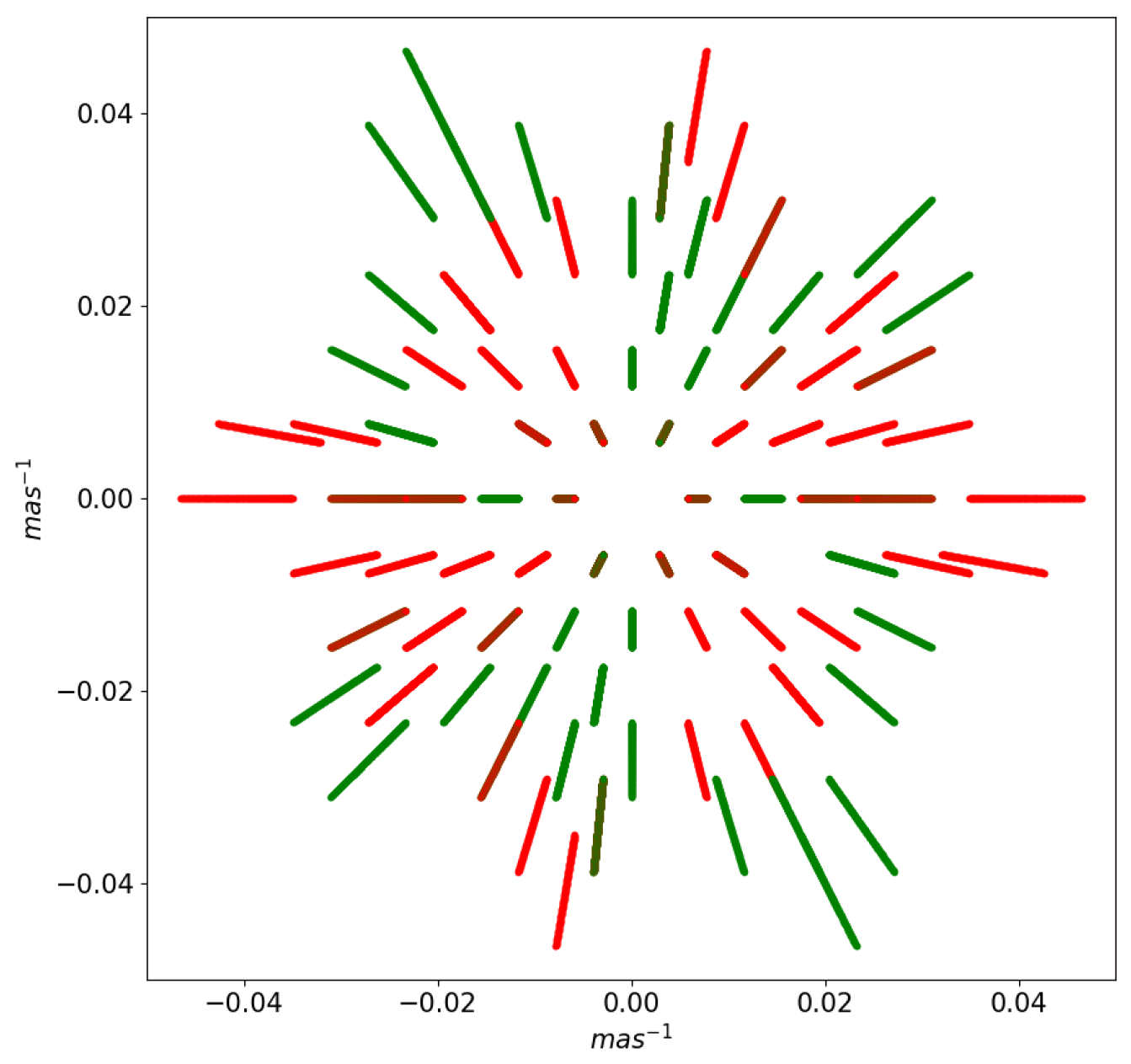} \\
    \end{tabular}{}
    \caption{Left: Input pupil configuration for the 2 sets of 9 fibers each. Right: (u,v) plane coverage of both configurations for a wavelengths ranging from $600nm$ to $800nm$. The colors of the lines match with the color of the left side configurations.}
    \label{FIRST_Subaru}
\end{figure}

The configuration of the two sets of fibers are showed in Fig.\ref{FIRST_Subaru}-left. The corresponding spatial frequency coverage is provided Fig.\ref{FIRST_Subaru}-right for wavelength ranging from $600nm$ to $800nm$. 

{}

\section{data reduction}
\label{sec-datared}

\subsection{The data reduction pipeline: extraction of the baseline complex coherence}

A typical FIRST image is composed of two sets of fringes (one of them showed Fig.~\ref{FIRST_images}-right - one per fiber set) that are analyzed independently but rely on the same fringe fitting technique, well suited for compact interferometer data analysis~\cite{tatulli2007interferometric}. The data reduction, extensively derived and described by \textit{Huby E.}\cite{huby2012first,huby2013caracterisation}, aims at retrieving the complex coherence terms ($\mu_{nn'}$) for each baseline ($n-n'$) from the fringe pattern, written as:
\begin{align}
	\mu_{nn'}=|V_{nn'}|\text{e}^{i\psi_{nn'}}A_nA_{n'}\text{e}^{i\Delta\Phi_{nn'}}.\label{mu}
\end{align}
$|V_{nn'}|$ and $\psi_{nn'}$ are respectively the object's complex visibility modulus and phase. $A_n$, $A_{n'}$ and $\Delta\Phi_{nn'}$ are respectively sub-pupils n and n' flux and differential piston. Considering N fibers providing N interfering beams, the fringe intensity can be written and developed, for each wavelength, as: 
\begin{align}
I(\Vx)= &\GP{\sum_{n<N}A_nE_n(\Vx)}^2
= \sum_{n<N}A_n^2E_n^2(\Vx) + 2 \text{Re}\GC{\sum_{n<n'<N} A_nE_n(\Vx)A_{n'}E_{n'}(\Vx)\text{e}^{i(2\pi f_{nn'}\Vx+\Delta\Phi_{nn'})}} \nonumber\\
= & \sum_{n<N}A_n^2E_n^2(\Vx) + 2 \sum_{n<n'<N} A_nE_n(\Vx)A_{n'}E_{n'}(\Vx)\text{cos}\GP{2\pi f_{nn'}\Vx+\Delta\Phi_{nn'}},\label{eq2}
\end{align}
with $\Vx$ the spatial variable (pixel index), $E_n$ the normalized envelope of each beam and $f_{nn'}$ the $nn'$ baseline frequency. By taking into account the object's complex visibility and re-arranging Eq.(\ref{eq2}), it is possible to obtain a linear relationship between the interferogram $I(\Vx)$ and the complex coherence (Eq.~\ref{mu}):
\begin{equation}
I(\Vx) = \mu_0E_g(\Vx)\label{Eq3} +\sum_{n<n'<N}\mathcal{R}\{\ \mu_{nn'}\}C_{nn'}(\Vx)+\sum_{n<n'<N}\mathcal{I}\{\ \mu_{nn'}\}S_{nn'}(\Vx)
\end{equation}
with $\mu_0$ the total flux, $Eg(\Vx)$ the normalized global envelope function and
\begin{align}
\left\{  
\begin{array}{r@{}l}\displaystyle
&C_{nn'}(\Vx)= 2E_n(\Vx)E_n'(\Vx)\cos(2\pi f_{nn'}\Vx)\nonumber\\
&S_{nn'}(\Vx)=-2E_n(\Vx)E_n'(\Vx)\sin(2\pi f_{nn'}\Vx)\nonumber
\end{array}\right. 
\end{align}
$\{ E_g(\Vx) \text{;} C_{nn'}(\Vx) \text{;} S_{nn'}(\Vx)  \}$ form a base on which the $I(\Vx)$ interferogram can be decomposed. This base depends on the instrument setup. Let us define $x_k$ the pixel index with $k \in \{1,...,n_p\} $ with $n_p$ the number of pixels. We also concatenate indexes $nn'$ into $\{1,...,n_b\}$ with $n_b$ the number of baseline pairs.  Finally the unknowns ($\mu_{n_b}$) can then be regrouped in a vector \textbf{P} and Eq.~(\ref{Eq3}) can be written as the matrix product:
\begin{align}
\text{\textbf{I}}=
\begin{bmatrix}
	I_{x_1} \\
	\vdots \\
	I_{x_{np}} 
\end{bmatrix}
= V2PM \cdot
\begin{bmatrix}
\mathcal{R}\{\ \mu_{1} \} \\
\vdots \\
\mathcal{R}\{\ \mu_{n_B} \} \\
\mathcal{I}\{\ \mu_{1} \} \\
\vdots \\
\mathcal{I}\{\ \mu_{n_B} \} \\
\mu_0
\end{bmatrix}
=V2PM\cdot\text{\textbf{P}}
\end{align}
with
\begin{align}
	V2PM=
	\begin{bmatrix}\label{v2pm}
	C_1(x_1) & \dots & C_{n_B}(x_1)  & S_1(x_1) & \dots & S_{n_B}(x_1) & E_g(x_1) \\
	\vdots &  & \vdots & \vdots &  & \vdots &   \vdots \\
	C_1(x_{n_p}) & \dots & C_{n_B}(x_{n_p})  & S_1(x_{n_p}) & \dots & S_{n_B}(x_{n_p}) & E_g(x_{n_p})
	\end{bmatrix}\nonumber
\end{align}
According to the formalism introduced by \textit{Millour et al.}~\cite{millour2004data}, the V2PM is the Visibility-To-Pixel matrix with a size of $(2n_b+1)\times n_p$. This matrix is rectangular hence it cannot be inverted. However, because the recombination on FIRST is non-redundant, each baseline frequency is unique hence the $V2PM$ modes are orthogonal one to another. In this case, we can compute $V2PM^\dag$, the partial generalized inverse of $V2PM$ matrix computed from the Singular Value Decomposition of $V2PM$. We can then compute the estimate $\EST{\textbf{P}}$ according to:
\begin{equation}\label{projection}
	\EST{\textbf{P}}=V2PM^\dag\cdot\text{\textbf{I}},
\end{equation}

which allows to reconstruct the baselines complex coherences.

\subsection{Closure phase measurements for resolved object characterization}

The main goal of the data reduction pipeline is to provide spatial and spectral information about resolved objects (\textit{e.g.} stellar binaries or giant stars). This information is contained in the object's complex visibilities. The way to extract the latter is through closure phase (CP) measurements, which are by definition the phase of the bispectrum~$\mu_{nn'n''}$:
\begin{equation}\label{bispectrum}
	\mu_{nn'n''}=<\mu_{nn'}\mu_{n'n''}\mu_{nn''}^*>,
\end{equation}
where $n$, $n'$ and $n''$ are the sub-pupil indexes used to form the triangle and $<>$ the average. The nice feature of this quantity is that it allows to cancel differential phase errors between sub-pupils since they cancel each other out according to Eq.(\ref{mu}):
\begin{equation}\label{bispectrum_angle}
	\text{CP}_{nn'n''}=\text{arg}(\mu_{nn'n''}) = \psi_{nn'}+\psi_{n'n''}-\psi_{nn''},
\end{equation}
The CP measurements are then fitted to a model to estimate physical parameters of the observed object. In this paper, we only focus on the case of a stellar binary. In this particular case, the object complex visibility, considering one of the component in the center of the field of view, is written as:
\begin{equation}\label{Vis_bin}
    V=\frac{1}{1+\rho}(1+\rho\text{e}^{-2i\pi\V{\Delta}\cdot \V{f}}),
\end{equation}
with $\rho$ the flux ratio between the two components, $\Delta$ the separation vector ($\alpha$,$\beta$) between the two components, and $\V{f}$ the spatial frequency of the considered baseline. This allows us to model the expected CP measurements for various flux ratios and geometries of binary systems. We then perform a fitting of the estimated CP ($CP_{estimate}$) with the model ($CP_{model}$). To do so, we want to minimize the $\chi^2$ function defined for each spectral channel as:
\begin{equation}
    \chi^2(\rho,\alpha,\beta) = \sum_{k}^{n_{CP}} \frac{(CP_{estimate}^k - CP_{model}^k(\rho,\alpha,\beta))^2}{{\sigma^{k}}^2}
\end{equation}
with $\sigma^{k}$ the error on the $k^{th}$ $CP_{estimate}$. We can then derive the likelihood function from $\chi^2$ as:
\begin{equation}\label{Like_fct}
    \mathcal{L}(\rho,\alpha,\beta) \propto \text{e}^{-\frac{\chi^2(\rho,\alpha,\beta)}{2}}
\end{equation}
Each parameter can then be retrieved by integrating $ \mathcal{L}$ over all the other parameters.

\section{On-sky results}

We present here preliminary on-sky results acquired on the Capella binary star on the $16^{th}$ of September 2020. This binary has an historic significance for FIRST, since it was the first binary studied by the instrument during its installation on the Lick observatory~\cite{huby2013first}. Capella has a magnitude of $-0.52$ in R-band, the expected separation of around $45mas$ between the two components, and their flux ratio is close to 1. Even though the separation is at least twice the diffraction limit of the telescope in the visible, this target is a great candidate to validate the ability of the FIRST instrument to detect stellar companionship at the Subaru telescope, and test our data reduction pipeline. The designated calibrator was rho Per, a star not resolved by the Subaru Telescope, with a magnitude of 1.59 in R-band.

\subsection{Direct imaging of Capella with SCExAO/VAMPIRES}

\begin{figure}[!h]
    \centering
    \includegraphics[width=0.23\linewidth]{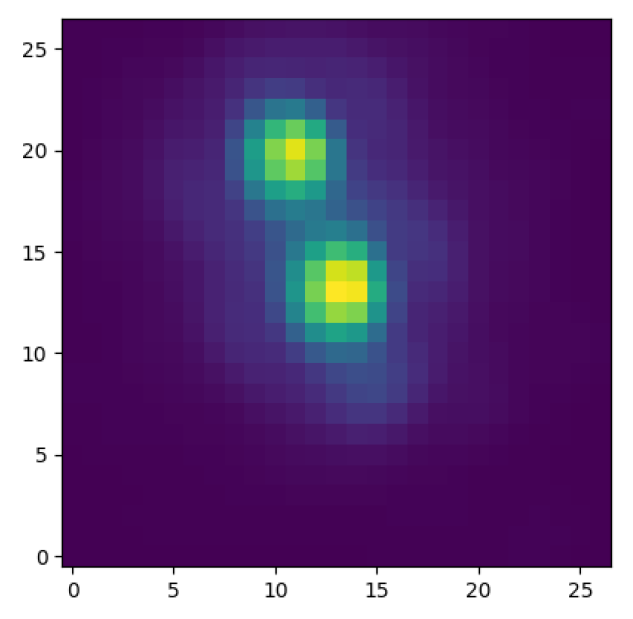}  \includegraphics[width=0.69\linewidth]{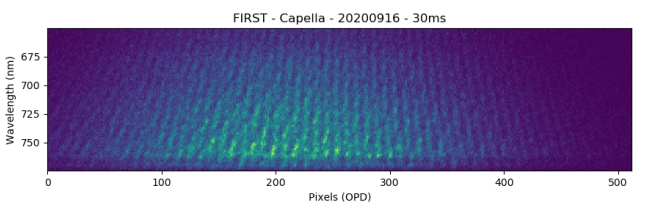} 
    \caption{Left: Capella binary system imaged with VAMPIRES. Estimated separation: $44\pm 1 mas$. Right: Fringe pattern acquired with the FIRST instrument.}
    \label{Capella_Vamp}
\end{figure}

Before presenting the results obtained with the FIRST instrument, we show here an image acquired with a focal plane camera on the SCExAO visible bench (Fig.~\ref{Capella_Vamp}-left) @ $750$~nm. This camera belongs to the VAMPIRES instrument, and shows the quality of the extreme adaptive optics system in the visible. The presented processed image is a stack of the best images selected from 30~data cubes, each containing $10000$~images. A basic method of image selection (based on the maximum intensity of the image) was used to select the best images of the data cubes. \textit{J. Lozi} developed a binary fitting tool to extract an estimation of the separation between the two components of the binary. The analyze of the Capella image result gives a separation of $44\pm 1$ mas. The plate scale of the image is $6.1$~mas per pixel, and the field showed here is about $160$mas, comparable to the FIRST field of view.\\

{}

\subsection{Capella binary detection with FIRST}

\begin{figure}[!h]
    \centering
    \begin{tabular}{cc}
        \includegraphics[width=0.455\linewidth]{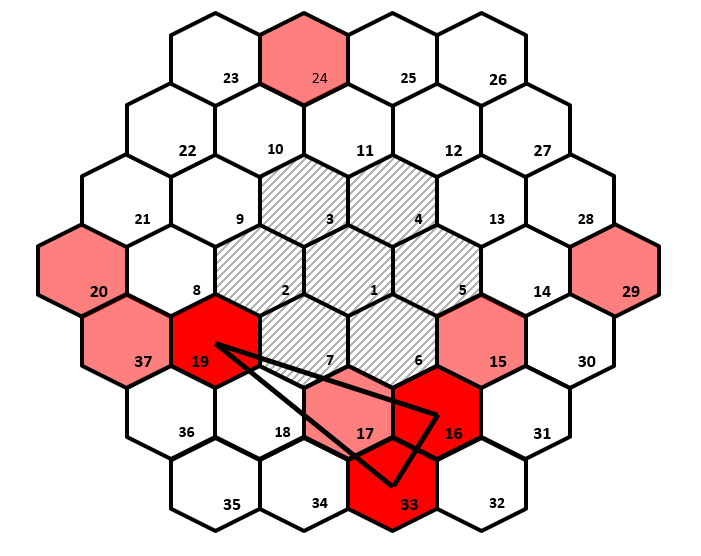} & \includegraphics[width=0.475\linewidth]{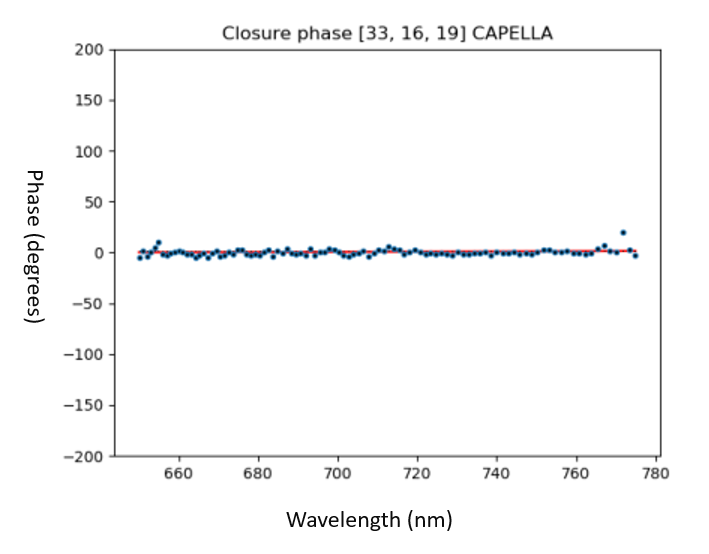} \\
    \end{tabular}{}
    \caption{Left: Selected triangle formed by sub-apertures corresponding to segments $[16,19,33]$ in the entrance pupil. Right: Closure phase signal (blue dots) extracted from the Capella data for the selected triangle. The best fit provided by the data reduction is over-plotted (red solid line).}
    \label{Capella_CP1}
\end{figure}

\begin{figure}[!h]
    \centering
    \begin{tabular}{cc}
        \includegraphics[width=0.455\linewidth]{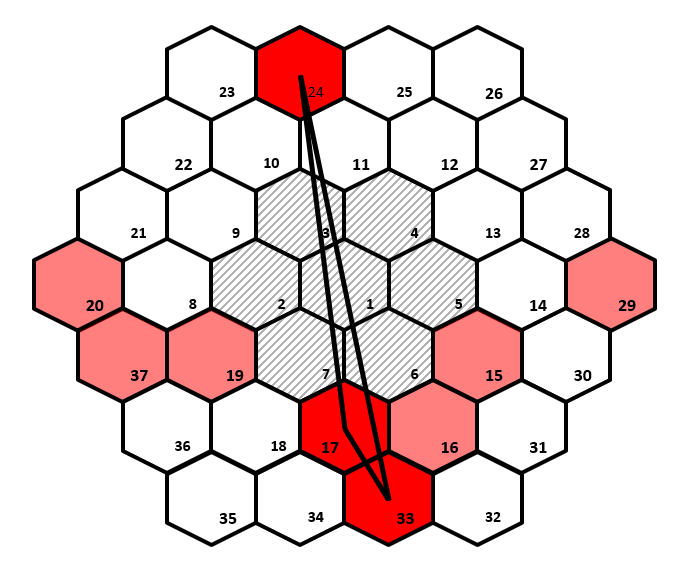} & \includegraphics[width=0.475\linewidth]{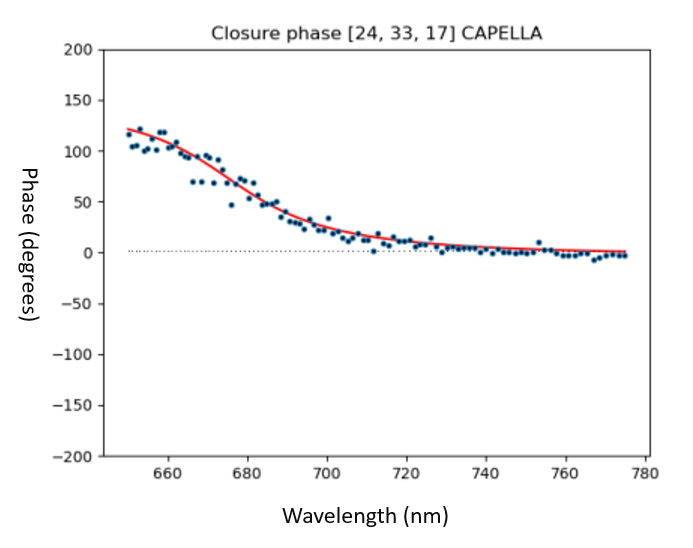} \\
    \end{tabular}{}
    \caption{Left: Selected triangle formed by sub-apertures corresponding to segments $[17,24,33]$ in the entrance pupil. Right: Closure phase signal (blue dots) extracted from the Capella data for the selected triangle. The best fit provided by the data reduction is over-plotted (red solid line).}
    \label{Capella_CP2}
\end{figure}

The data was acquired with only one of the two sets of fibers (the red on Fig.~\ref{FIRST_Subaru}) for technical issues, and with disregarding the sub-aperture conjugated with segment 15 because of the deterioration of the Iris AO MEMS. We used the data reduction pipeline described in Section~\ref{sec-datared} to try and retrieve the parameters (contrast, separation) of the Capella binary out of the $28$~baselines interference signal. A cube of 500 frames with a $30$ms exposure time was used (one frame is showed Fig.~\ref{Capella_Vamp}-right). Using Eqs.~(\ref{Vis_bin}-\ref{Like_fct}) we fit the 56 extracted closure phase measurements. The point source Target Rho Per was used as a calibrator, to remove the instrumental bias induced in the closure phase measurements. We show the resulting calibrated closure phase measurements for two selected triangles on Figures~\ref{Capella_CP1}~and~\ref{Capella_CP2}. 

{}

The triangle showed Fig.~\ref{Capella_CP1} is formed by the sub-apertures conjugated with segments $[16,19,33]$ of the Iris AO MEMS. We can see on the Fig.~\ref{Capella_CP1}-right the on-sky closure phase measurement (blue dots) overlapped with the best fit obtained (solid red line). The average value of the calibrated closure phase here is $0.1 \pm 3.2 \deg$, meaning that the observed object was not resolved by any of these baselines. The remaining uncalibrated bias generating the rather large standard deviation results from instrumental effect that are still to be understood and tackled down. The average of the closure phase error bars is $1.2\deg$.

{}

The triangle showed Fig.~\ref{Capella_CP2} is formed by the sub-apertures conjugated with segments $[17,24,33]$ of the Iris AO MEMS. In this case, as we can see on the obtained calibrated closure phase measurement (blue dots Fig.~\ref{Capella_CP2}~right), the Capella binary system is detected - a non-zero signal appears in the measurement. In this case, the baselines constituting the triangle have higher spatial frequencies compared to the previous ones, therefore it allows to resolve the binary system. Usually, for a symmetric binary, the detected signal is a phase shift between $\pm 180 \deg$ and $0\deg$. Here, we can see a not-so-abrupt phase shift from $125\deg$ to $0\deg$. This means that the detected binary should be asymmetric, meaning that the flux ratio between the two components is below~1. The average of the closure phase error bars is $1.5\deg$. Over-plotted on the graph (red solid line), we can see the result of the closure phase fitting measurement, which we will discuss now.

\begin{figure}[!h]
    \centering
    \includegraphics[width=0.8\linewidth]{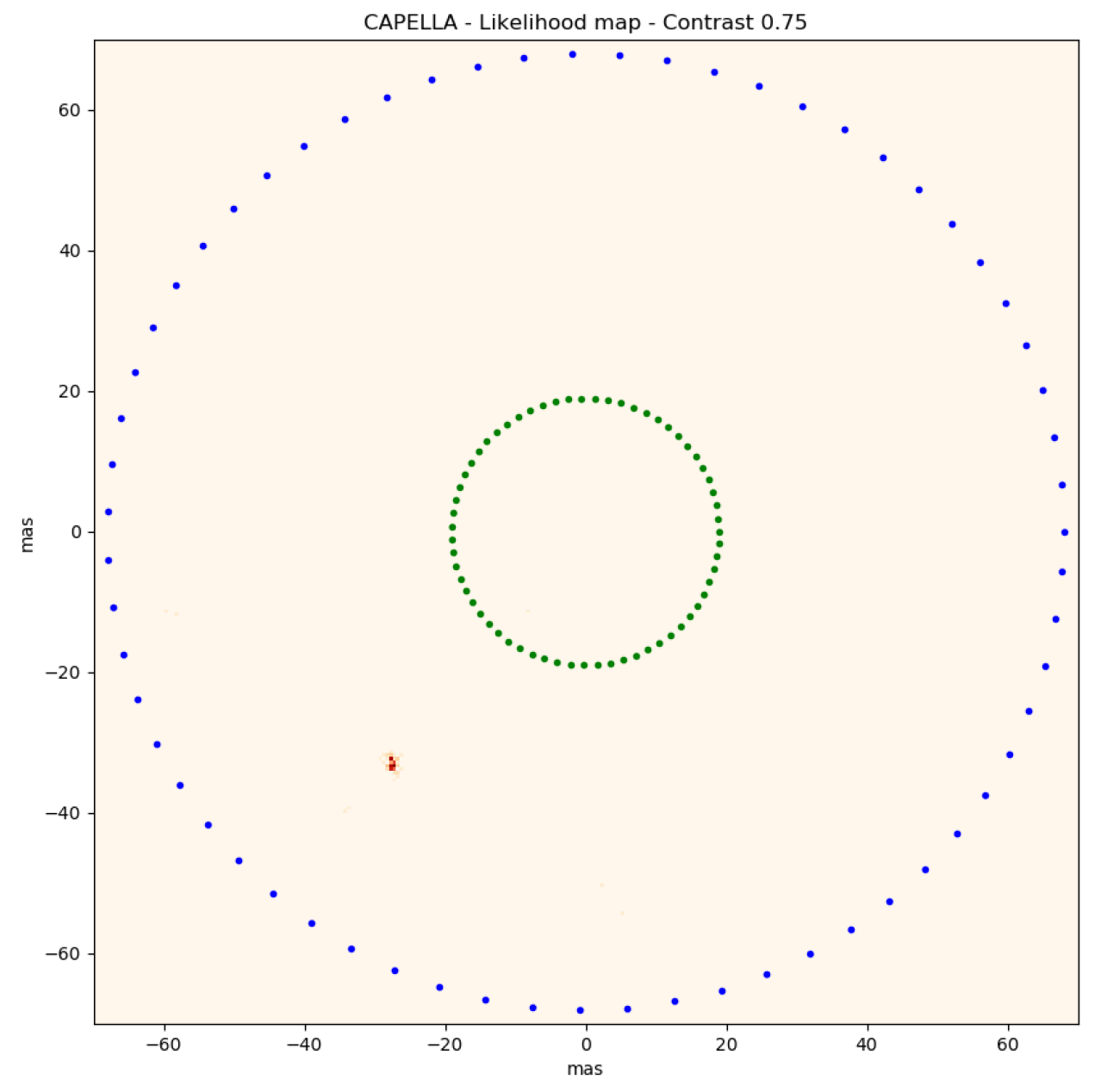}
    \caption{Likelihood map in the case of Capella. The green and blue circles indicate respectively the telescope's diffraction limit @$750nm$ ($19$mas) and the FIRST instrument field of view ($136$mas) at $700$nm. The blob located at $[-27.5,-33]$ mas shows the estimated position of the companion in the Capella binary system, $43.3$ mas away from the central star. No absolute calibration of the position has been done yet.}
    \label{Capella_Lmap}
\end{figure}

Fig.~\ref{Capella_Lmap} shows the likelihood map delivered by the data reduction (see Eq.~\ref{Like_fct}). The closure phase model fitting was performed on a 140 by 140 mas window, with a $0.5$~mas step. The green circle shows the telescope diffraction limit @~$750$~nm (19 mas). The blue circle indicates FIRST field of view (136 mas). The blob that can be seen at coordinates $[-27.5,-33]$ mas corresponds to the estimated position of the second component of the binary (the first component being in the center of the map), where the fitting of the closure phase model is the best. The estimated separation provided by the current data reduction process gives $43.3 \pm 0.7$ mas. This value is in accordance with the estimated separation of the components with direct imaging ($44\pm 1$ mas). The error bar on our estimation should be minimized after more precise statistical analysis of our closure phase fitting test (frame selection, aberrant values filtering, fitted closure phase selection...), but still offers a more accurate precision compared to direct imaging. The average estimated contrast ratio between the two components here is $0.75$ (a precise estimation of the flux ratio as the function of the wavelength will be done in the near future). This is lower than expected, and can be due to a poorer injection of the off-axis companion into the single mode fiber, compared to the central star. This needs to be carefully calibrated, along with the plate scale. 

{}

\subsection{Discussion}

To better characterize the instrument, we plan on using what we call a "fake binary". The goal here will be to apply fast tip/tilt on the SCExAO deformable mirror (at about 1kHz), generating 2 incoherent sources, and acquire data with FIRST (at about 1Hz). The tip/tilt amplitude will determine the separation of this "fake binary" and the ratio of applied tips and tilts can determine the contrast ratio between the two components (for example, if we apply twice the amount of tips compared to the tilts, the contrast ratio will be 0.5). This will allow us to determine the precise plate scale of the instrument, and characterize the injection losses due to off-axis sources.

\section{conclusion}

We presented the FIRST spectro-interferometer instrument, integrated as a module of SCExAO on the Subaru Telescope, and benefiting from a high quality and stable wavefront in the Visible. We showed that the preliminary analysis of on-sky data allowed to detect the Capella binary system, with a $43.3 \pm 0.7$ mas separation and a contrast ratio of $0.75$ on average. This separation value is in accordance with direct imaging of the same system performed almost simultaneously. With more work on the data reduction, the FIRST instrument will hopefully bring more exciting science results in the near future. Some characterization work still needs to be done, for example with the "fake binary" tests for plate-scale measurements and off-axis sources injection losses assessment. Moreover, new upgrades are currently under development~\cite{BarjotSPIE,MartinSPIE} to better the precision of the phase measurement. These new upgrades will be installed in the near future, on one of the fiber sets (while the other can continue being exploited on-sky). Finally, we are exploring the possibility to use FIRST as an interferometric  wavefront sensor with spectral resolution. This would help tackling down aberrations originating from pupil fragmentation, and would make FIRST a powerful science instrument with wavefront sensing and control capabilities.

\acknowledgments 
The development of FIRST was supported by Centre National de la Recherche Scientifique CNRS (Grant ERC LITHIUM - STG - 639248). The development of SCExAO was supported by the Japan Society for the Promotion of Science (Grant-in-Aid for Research \#23340051, \#26220704, \#23103002, \#19H00703 \& \#19H00695), the Astrobiology Center of the National Institutes of Natural Sciences, Japan, the Mt Cuba Foundation and the director's contingency fund at Subaru Telescope. The authors wish to recognize and acknowledge the very significant cultural role and reverence that the summit of Maunakea has always had within the indigenous Hawaiian community. We are most fortunate to have the opportunity to conduct observations from this mountain.

\bibliography{main} 

\begin{thebibliography}{10}

\bibitem{rousset1990first}
Rousset, G., Fontanella, J., Kern, P., Gigan, P., and Rigaut, F., ``First
  diffraction-limited astronomical images with adaptive optics,'' {\em
  Astronomy and Astrophysics}~{\bf 230},  L29--L32 (1990).

\bibitem{labeyrie1970attainment}
Labeyrie, A., ``Attainment of diffraction limited resolution in large
  telescopes by fourier analysing speckle patterns in star images,'' {\em
  Astron. Astrophys.}~{\bf 6}(1),  85--87 (1970).

\bibitem{lacour2011sparse}
Lacour, S., Tuthill, P., Amico, P., Ireland, M., Ehrenreich, D., Huelamo, N.,
  and Lagrange, A.-M., ``Sparse aperture masking at the vlt-i. faint companion
  detection limits for the two debris disk stars hd 92945 and hd 141569,'' {\em
  Astronomy \& Astrophysics}~{\bf 532},  A72 (2011).

\bibitem{perrin2006high}
Perrin, G., Lacour, S., Woillez, J., and Thi{\'e}baut, E., ``High dynamic range
  imaging by pupil single-mode filtering and remapping,'' {\em Monthly Notices
  of the Royal Astronomical Society}~{\bf 373}(2),  747--751 (2006).

\bibitem{2015PASP..127..890J}
{Jovanovic}, N., {Martinache}, F., {Guyon}, O., {Clergeon}, C., {Singh}, G.,
  {Kudo}, T., {Garrel}, V., {Newman}, K., {Doughty}, D., {Lozi}, J., {Males},
  J., {Minowa}, Y., {Hayano}, Y., {Takato}, N., {Morino}, J., {Kuhn}, J.,
  {Serabyn}, E., {Norris}, B., {Tuthill}, P., {Schworer}, G., {Stewart}, P.,
  {Close}, L., {Huby}, E., {Perrin}, G., {Lacour}, S., {Gauchet}, L.,
  {Vievard}, S., {Murakami}, N., {Oshiyama}, F., {Baba}, N., {Matsuo}, T.,
  {Nishikawa}, J., {Tamura}, M., {Lai}, O., {Marchis}, F., {Duchene}, G.,
  {Kotani}, T., and {Woillez}, J., ``{The Subaru Coronagraphic Extreme Adaptive
  Optics System: Enabling High-Contrast Imaging on Solar-System Scales},'' {\em
  Publications of the Astronomical Society of the Pacific}~{\bf 127},  890
  (Sept. 2015).

\bibitem{vievard2020capabilities}
Vievard, S., Cvetojevic, N., Huby, E., Lacour, S., Martin, G., Guyon, O., Lozi,
  J., Kotani, T., Jovanovic, N., Perrin, G., et~al., ``Capabilities of a
  fibered imager on an extremely large telescope,'' {\em arXiv preprint
  arXiv:2010.10733}  (2020).

\bibitem{huby2012first}
Huby, E., Perrin, G., Marchis, F., Lacour, S., Kotani, T., Duch{\^e}ne, G.,
  Choquet, E., Gates, E., Woillez, J., Lai, O., et~al., ``First, a fibered
  aperture masking instrument-i. first on-sky test results,'' {\em Astronomy \&
  Astrophysics}~{\bf 541},  A55 (2012).

\bibitem{huby2013first}
Huby, E., Duch{\^e}ne, G., Marchis, F., Lacour, S., Perrin, G., Kotani, T.,
  Choquet, {\'E}., Gates, E., Lai, O., and Allard, F., ``First, a fibered
  aperture masking instrument-ii. spectroscopy of the capella binary system at
  the diffraction limit,'' {\em Astronomy \& Astrophysics}~{\bf 560},  A113
  (2013).

\bibitem{kotani2009pupil}
Kotani, T., Lacour, S., Perrin, G., Robertson, G., and Tuthill, P., ``Pupil
  remapping for high contrast astronomy: results from an optical testbed,''
  {\em Optics Express}~{\bf 17}(3),  1925--1934 (2009).

\bibitem{minowa2010performance}
Minowa, Y., Hayano, Y., Oya, S., Watanabe, M., Hattori, M., Guyon, O., Egner,
  S., Saito, Y., Ito, M., Takami, H., et~al., ``Performance of subaru adaptive
  optics system ao188,'' in [{\em Adaptive Optics Systems
  II}{\nolinebreak\hspace{0.1em}]},   {\bf 7736},  77363N, International
  Society for Optics and Photonics (2010).

\bibitem{Lozi_2019}
Lozi, J., Jovanovic, N., Guyon, O., Chun, M., Jacobson, S., Goebel, S., and
  Martinache, F., ``Visible and near-infrared laboratory demonstration of a
  simplified pyramid wavefront sensor,'' {\em Publications of the Astronomical
  Society of the Pacific}~{\bf 131},  044503 (mar 2019).

\bibitem{Currie_2018}
Currie, T., Brandt, T.~D., Uyama, T., Nielsen, E.~L., Blunt, S., Guyon, O.,
  Tamura, M., Marois, C., Mede, K., Kuzuhara, M., Groff, T.~D., Jovanovic, N.,
  Kasdin, N.~J., Lozi, J., Hodapp, K., Chilcote, J., Carson, J., Martinache,
  F., Goebel, S., Grady, C., McElwain, M., Akiyama, E., Asensio-Torres, R.,
  Hayashi, M., Janson, M., Knapp, G.~R., Kwon, J., Nishikawa, J., Oh, D.,
  Schlieder, J., Serabyn, E., Sitko, M., and Skaf, N., ``{SCExAO}/{CHARIS}
  near-infrared direct imaging, spectroscopy, and forward-modeling of kappa and
  b: A likely young, low-gravity superjovian companion,'' {\em The Astronomical
  Journal}~{\bf 156},  291 (nov 2018).

\bibitem{vampires}
Norris, B., Schworer, G., Tuthill, P., Jovanovic, N., Guyon, O., Stewart, P.,
  and Martinache, F., ``{The VAMPIRES instrument: imaging the innermost regions
  of protoplanetary discs with polarimetric interferometry},'' {\em Monthly
  Notices of the Royal Astronomical Society}~{\bf 447},  2894--2906 (01 2015).

\bibitem{jovanovic2017developing}
Jovanovic, N., Guyon, O., Kotani, T., Kawahara, H., Hosokawa, K., Lozi, J.,
  Males, J., Ireland, M., Tamura, M., Mawet, D., et~al., ``Developing
  post-coronagraphic, high-resolution spectroscopy for terrestrial planet
  characterization on elts,'' {\em arXiv preprint arXiv:1712.07762}  (2017).

\bibitem{huby2013caracterisation}
Huby, E., {\em Caract{\'e}risation de syst{\`e}mes binaires par imagerie haute
  dynamique non redondante fibr{\'e}e}, PhD thesis, Observatoire de Paris
  (2013).

\bibitem{tatulli2007interferometric}
Tatulli, E., Millour, F., Chelli, A., Duvert, G., Acke, B., Utrera, O.~H.,
  Hofmann, K.-H., Kraus, S., Malbet, F., M{\`e}ge, P., et~al.,
  ``Interferometric data reduction with amber/vlti. principle, estimators, and
  illustration,'' {\em Astronomy \& Astrophysics}~{\bf 464}(1),  29--42 (2007).

\bibitem{millour2004data}
Millour, F., Tatulli, E., Chelli, A.~E., Duvert, G., Zins, G., Acke, B., and
  Malbet, F., ``Data reduction for the amber instrument,'' in [{\em New
  Frontiers in Stellar Interferometry}{\nolinebreak\hspace{0.1em}]},   {\bf
  5491},  1222--1230, International Society for Optics and Photonics (2004).

\bibitem{BarjotSPIE}
Barjot, K., Huby, E., Vievard, S., Cvetojevic, N., Lacour, S., Martin, G.,
  Guyon, O., Lozi, J., Jovanovic, N., Perrin, G.~S., Marchis, F., Kotani, T.,
  Lapeyrère, V., and Rouan, D., ``Laboratory characterization of firstv2
  photonic for the study of substellar companions,'' {\em Proceeding SPIE,
  Astronomical Telescopes and Instrumentation} ,  Paper 11446--48 (2020).

\bibitem{MartinSPIE}
Martin, G., Foin, M., Gardillou, F., Cassagnettes, F., Ulliac, G., Courjal, N.,
  Barjot, K., Cvetojevic, N., Vievard, S., Lapeyrère, V., Huby, E., and
  Lacour, S., ``Recent results on electro-optic visible multi-telescope beam
  combiner for next generation first/subaru instruments: Hybrid and 3d
  devices,'' {\em Proceeding SPIE, Astronomical Telescopes and Instrumentation}
  ,  Paper 11446--51 (2020).

\end{thebibliography}
\bibliographystyle{spiebib} 

\end{document}